\documentstyle{article}

\def\tilde{\widetilde}
\def\bar{\overline}

\def\*{\star}
\def\({\left(}          
\def\){\right)}         
\def\[{\left[}          
\def\]{\right]}

\def\frac#1#2{{#1 \over #2}}

\def\2pi{\hbox{$2\pi i$}}

\def\dsl{\raise.15ex\hbox{/}\kern-.57em\partial}
\def\Dsl{\,\raise.15ex\hbox{/}\mkern-.13.5mu D}
\def\b{\beta}
\def\al{\alpha}

              \def\CO{{\cal O}}

\def\debut{ \begin{eqnarray} }
\def\fin{ \end{eqnarray} }
\def\non{ \nonumber }

\def\presentation{
\voffset -.50in
\hoffset -.19in
\oddsidemargin 0in \evensidemargin 0in
\marginparwidth .75in \marginparsep 7pt \topmargin 0in
\headheight 12pt \headsep .25in
\footheight 18pt \footskip .35in
\textheight 9.5in \textwidth 6.5in
\columnsep 10pt \columnseprule 0pt }

\presentation

\begin{document}

\rightline{LPTHE-96-36}
\vskip 1cm
\centerline{\LARGE Form factors for principal chiral field model}
\bigskip
\centerline{\LARGE with Wess-Zumino-Novikov-Witten term.}
\vskip 2cm
\centerline{\large \framebox{P. Mejean} and F.A. Smirnov
\footnote[1]{On leave from Steklov Mathematical Institute,
Fontanka 27, St. Petersburg, 191011, Russia} }
\vskip1cm
\centerline{ Laboratoire de Physique Th\'eorique et Hautes
Energies \footnote[2]{\it Laboratoire associ\'e au CNRS.}}
\centerline{ Universit\'e Pierre et Marie Curie, Tour 16 1$^{er}$
		\'etage, 4 place Jussieu}
\centerline{75252 Paris cedex 05-France}
 \vskip2cm
{\bf Abstract.} We construct the form factors of the trace
of energy-momentum tensor for the massless model described
by $SU(2)$ principal chiral field model with WZNW tern on
level 1. We explain how this construction can be generalized to
a class of integrable massless models including the flow
from tricritical to critical Ising model.

\newpage
\hskip 3cm
{\it From F. Smirnov.} During several months I worked with

\hskip 3cm Pierre Mejean. After his premature decease which deeply

\hskip 3cm affected everybody who new him I decided to collect and

\hskip 3cm to publish the results which we obtained together.

\section{Introduction}

There is a wide class of integrable models which describe the
flows between two different models of Conformal Field Theory (CFT)
in UV and IR regions \cite{zam}. To describe these models the factorized
scattering of massless particles was proposed in \cite{zz1}. In spite
of general difficulties arising from the fact that the scattering
of massless particles is not properly defined in 2D the application
of this method provided very good results for several particular models.
Certainly, the physical results are related to possibility of
extracting the off-shell information from these S-matrices.
Similarly to the massive case there are two ways to do that.

The first way is TBA approach. The TBA equations can be written for
the massless scattering which allow to calculate the effective
central charge and to show that the latter really interpolates between
corresponding UV and IR values \cite{zz1}.

The second way consists in generalization proposed in \cite{muss}
of the form factor bootstrap
approach (which is originally formulated for the massive models \cite{book})
to the massless flows. One must distinguish between the form factor bootstrap
in massive and massless cases. In the massive case the form factor
bootstrap stays on the solid ground because the space of states
is well defined as the Fock space of particles. In the massless case
this very definition is doubtful and one has to consider the
form factor bootstrap rather as intuitive than rigorous
procedure. Indeed for certain operators (as, for example,
the order-disorder operators for the flow from tricritical to
critical Ising model) the straightforward application of the
method leads to divergent series for the Green functions \cite{muss}.
However, the situation can be improved even for those operators,
furthermore there are operators for which the series converge
providing spectacular examples of interpolation between UV and IR
limits \cite{muss}. So, the form factor bootstrap is a method
which works in massless case in spite of all the problems of
general character.

It must be noticed, however, that the complete construction
of all the form factors is not given in the paper \cite{muss}
even for the simplest model describing the
flow from tricritical to
critical Ising model
considered there. In the present paper
we shall give the complete construction for more complicated model:
the Principal Chiral Field with
Wess-Zumino-Novikov-Witten term.
We shall explain briefly that our construction can be generalized
to a wide class of models including the one considered in \cite{muss}.

\section{Formulation of the problem.}

On of the most beautiful examples of the massless flows for
which the S-matrix is known is given by the Principal Chiral Field with
Wess-Zumino-Novikov-Witten (WZNW) term on level $1$ (PCM${}_1$):
\debut
S={1\over 2\lambda ^2}\int tr(g^{-1}\partial_{\mu}g)
(g^{-1}\partial_{\mu}g) d^2x\ +\ i\Gamma (g)
\non
\fin
where the WZNW term $\Gamma (g)$ is defined by means of continuation
of $g$ to 3D manifold $B$ for which the 2D space-time is the boundary:
\debut
\Gamma (g)=\int\limits _{B} \epsilon ^{\mu \nu \lambda}
tr(g^{-1}\partial_{\mu}g)
(g^{-1}\partial_{\nu}g)(g^{-1}\partial_{\lambda}g)d^3x
\non
\fin
In the UV limit the pure PCM action dominates the central charge
being equal to 3. In the IR region the flow is attracted
by the fixed point $\lambda ^2=8\pi$ which corresponds to the conformal
WZNW model with the central charge equal to 1. It is explained in \cite{zz1}
that the flow arrives at the IR point along the direction
defined by the irrelevant operator $\bar{T}T$ composed of the right
and left components of the energy-momentum tensor.

In the IR region the theory is conformal, so, two charialities
essentially decouples. One describes corresponding left and right
level-1 WZNW models in terms of massless particles.
This is exactly left-left and right-right scattering which
seems to be doubtful in 2D. The prescription of the
paper \cite{zz1} for the definition of this scattering can be
understood in the following way.
We know that the level-1 WZNW model coincides with the UV limit of
the massive $SU(2)$-invariant Thirring model.
The local operators for the latter model are defined
via their form factors \cite{book}.
For the operators
chiral in the limit (as chiral components of the energy-momentum
tensor) the limiting values of the correlation functions
are obtained by replacing the massive dispersion
by the massless ones.
On the other hand these
limiting correlation functions
coincide with the conformal ones. So, we get the representation
of the conformal correlation functions for chiral operators in terms of
the form factor series with massless particles.
The form factors are defined through the S-matrix for the original massive
theory which is considered now as S-matrix of massless
particles.
It is proposed to
use this representations
of the correlators as a starting
point of the description of massless flows.

More precisely, left
and right particles are parametrized by the rapidities
$\b$ and $\b'$ such that the energy-momentum are respectively
$$e=-p=Me^{-\b},\qquad  e=p=Me^{\b'} $$
where $M$ is the mass scale which can be chosen arbitrary
on this stage. The theory possesses $SU(2)_L\otimes SU(2)_R$
symmetry, the left (right) movers are doublets with respect
to $SU(2)_L$ ($SU(2)_R$). The factorizable S-matrices which
describe the left and right CFT are given by
\debut
S_{LL}(\b _1,\b_2)=S^Y(\b _1-\b_2),\qquad
S_{RR}(\b _1',\b_2')=S^Y(\b _1'-\b_2')
\non
\fin
where $S^Y(\b)$ is the Yangian S-matrix for the scattering
of spin-${1\over 2}$ particles \cite{zz}:
\debut
S^Y(\b)={\Gamma ({1\over2}+{\b\over 2\pi i})\Gamma (-{\b\over 2\pi i})\over
\Gamma ({1\over2}-{\b\over 2\pi i})\Gamma ({\b\over 2\pi i}) }
\({\b I-\pi i P\over \b-\pi i}\)
\non\fin
where $I$ and $P$ are respectively unit and permutation
operators acting in the tensor product of two 2-dimensional
isotopic spaces.

The crucial point is in introducing the non-trivial left-right and right-left
S-matrices. Contrary to $S_{LL}$ and $S_{RR}$ whose definition is rather formal
the S-matrices $S_{LR}$ and $S_{RL}$
allow quite rigorous interpretation.
For the PCM${}_1$ the proposal of \cite{zz1} is
$$S_{RL}(\b'-\b)={1\over S_{LR}(\b-\b')}=U(\b'-\b),\qquad
U(\b)=\tanh{1\over2}\(\b -{\pi i\over 2}\) $$
The scale normalization $M$ is fixed by the requirement that
the zero of this S-matrix is situated exactly
at $\b ={\pi i\over 2}$.
It is quite amusing that the IR limit corresponds to
$\b-\b '=\log \Lambda$, $\Lambda\to\infty$, indeed in this
limit the $s$-variable goes to zero. This fact is very
interesting because in the massive case infinite rapidities are
always related to UV behaviour of the form factors which has
been investigated in several cases in \cite{book}, so, we can
use the familiar methods for solving absolutely different problems.

Let us describe the form factor bootstrap approach to
massless flows as it is formulated in \cite{muss}.
Consider the matrix element of certain local operator
$\CO$ taken between the vacuum and the state containing
The left  and right particles with rapidities
$\b_1,\cdots,\b_l$ and $\b_1',\cdots,\b_k'$ respectively:
\debut
f_{\CO}(\b_1,\cdots,\b_l\ |\ \b_1',\cdots,\b_k')
\label{Fb}
\fin
It is very convenient to collect all the rapidities together
into the set $\theta _1,\cdots,\theta _{k+l}=$
$\b_1,\cdots,\b_l,\b_1',\cdots,\b_k'$ and to introduce index
$a_i=L,R$ which distinguish the left and right particles.
The first requirement of the form factors is that
of symmetry:
\debut
f_{\CO}(\cdots ,\theta _i,\theta _{i+1},\cdots)_{\cdots,a_i,a_{i+1},\cdots}
S(\theta _i-\theta _{i+1})_{a_i,a_{i+1}}  =
f_{\CO}(\cdots ,\theta _{i+1},\theta _i,\cdots)_{\cdots,a_{i+1},a_i,\cdots}
\label{a1}
\fin
If $a_i\ne a_j$ this equation has to be considered as definition
which allows to construct the form factor with arbitrary placed
left and right particles starting from (\ref{Fb}).

The second requirement is
\debut
&&f_{\CO}(\theta _1,\cdots ,\theta _{k+l-1},\theta _{k+l}+2\pi i)
_{a_1,\cdots ,a_{k+l-1},a_{k+l}} =
f_{\CO}(\theta _{k+l},\theta _1,\cdots ,\theta _{k+l-1})
_{a_{k+l},a_1,\cdots ,a_{k+l-1}}=\non\\&&=
f_{\CO}(\theta _1,\cdots ,\theta _{k+l-1},\theta _{k+l})
_{a_1,\cdots ,a_{k+l-1},a_{k+l}}
S_{a_{k+l-1},a_{k+l}}(\theta _{k+l-1}-\theta _{k+l})
\cdots
S_{a_{1},a_{k+l}}(\theta _{1}-\theta _{k+l})
\label{a2}
\fin

Since we do not have bound states in the theory the form factor
$f_{\CO}(\theta _1,\cdots ,\theta _{k+l})$ is supposed to be
a meromorphic function of $\theta _{k+l} $ in the strip
$0<\theta _{k+l} <2\pi$ whose only singularities are the
simple poles at the points $\theta _{k+l} =\theta _{j} +\pi i$.
It is important that these singularities appear only in
left-left and right-right chanels. The residue at
$\theta _{k+l} =\theta _{k+l-1} +\pi i $
is given
by
\debut
&&2\pi i\ res
f_{\CO}(\theta _1,\cdots ,\theta _{k+l-2},\theta _{k+l-1},\theta _{k+l})
_{a_1,\cdots ,a_{k+l-2} ,a_{k+l-1},a_{k+l}}=
\non\\&&\hskip 1cm =
\delta _{a_{k+l-1},a_{k+l}}f_{\CO}(\theta _1,\cdots ,\theta _{k+l-2})
_{a_1,\cdots ,a_{k+l-2}}\otimes c_{k+l-1,k+l}
\non\\&&\hskip 1cm
\times
\(1\ -\ S_{a_{k+l-1},a_{1}}(\theta _{k+l-1}-\theta _{1})\cdots
S_{a_{k+l-1},a_{k+l-2}}(\theta _{k+l-1}-\theta _{k+l-2})\)
\label{a3}
\fin
here $c_{k+l-1,k+l}$ is a vector in the tensor product
of two isotopic spaces constructed from the charge conjugation
matrix, in our case it is the singlet vector in the tensor
product of two spin-${1\over 2}$ representations of $SU(2)$.

These requirements on the massless form factors do not differ too
much from the form factor axioms of \cite{book}. However, the
physical situation is quite different and the solutions to these
equation can not be found in \cite{book}.

\section{Form factors of the energy-momentum tensor.}

It the present paper we are going to construct the form factors
of the trace of energy-momentum tensor ($\Theta$) for PCM${}_1$.
Our methods are applicable to other operators, but we are
considering this particular one because of its nice properties
and physical importance.
Since the symmetry under $SU(2)_L\otimes SU(2)_R$ is not broken by
the perturbation the form factors have to be singlets with
respect to both isotopic groups. That is why $l=2n$ and $k=2m$.
The form factors of $\Theta$ satisfy general conditions (\ref{a1},
\ref{a2}, \ref{a3})
and additional requirements following from the fact that we consider
this particular operator.

1. The energy-momentum conservation implies that
\debut
f_{\Theta}(\b_1,\cdots,\b_{2n}\ |\ \b_1',\cdots,\b_{2m}')=
(\sum e^{-\beta _j})(\sum e^{\beta _j '})
f(\b_1,\cdots,\b_{2n}\ |\ \b_1',\cdots,\b_{2m}')
\label{c1}
\fin
where for $n>1$ and $m>1$ the function $f$ does not other
singularities that those of $f_{\Theta}$, for $n=1$($m=1$)
it has simple poles at $\b _2=\b _1 +\pi i$ ($\b _2 '=\b _1  '+\pi i$)
which are cancelled by $e^{-\b _1 } +e^{-\b _2 }$
($e^{\b _1' } +e^{\b _2 '}$ ).

2. The lowest form factor of $\Theta$ is that corresponding to
2+2 particles. However by the conservation law we can construct
from $f_{\Theta}$ the form factors of the left and right components of
the energy momentum tensor $T$ and $\bar{T}$ whose lowest
form factors are of the type $2n+0$ and $0+2m$ respectively.
These lowest form factors must coincide with the form
factors of pure $k=1$ WZNW model i.e. with those of
$SU(2)$-invariant Thirring model. One easily finds that it implies:
\debut
&&2\pi i\ res _{\b_2'=\b _1'+\pi i}\sum e^{-\b _j}\
f(\b_1,\cdots,\b_{2n}\ |\ \b_1',\b_{2}') =\non\\&&=
\widehat{f}_{T}(\b_1,\cdots,\b_{2n})\otimes c_{1',2'} \(1-
S_{LR}(\b _1'-\b_1)\cdots S_{LR}(\b _1'-\b_1)\) ,\non\\ &&
2\pi i \ res _{\b_2=\b _1+\pi i}\sum e^{\b _j'}\
f(\b_1,\b_{2}\ |\ \b_1',\cdots,\b_{2m}')=\non\\&&=
\widehat{f}_{\bar{T}}(\b_1',\cdots,\b_{2m}')
\otimes c_{1',2'} \(1-
S_{RL}(\b _1-\b_1')\cdots S_{RL}(\b _1-\b_1')\) \label{c2}
\fin
where $\widehat{f}_{T}$ and $\widehat{f}_{\bar{T}}$ are the form factors
of left and right components
of the energy-momentum tensor for the $SU(2)$ -Thirring model \cite{book}.

3. The IR limit corresponds to $\b _i-\b _j '\simeq\log\Lambda$ and
$\Lambda\to\infty$. In this limit one has to reproduce the
operator $T\bar{T}$ which defines the direction of the flow in the IR
region. So, we must have
\debut
f_{\Theta}(\b_1+\log\Lambda,\cdots,\b_{2n}+\log\Lambda|\ \b_1',\cdots,\b_{2m}')
\to
(M\Lambda) ^{-2 }
\widehat{f}_{T}(\b_1,\cdots,\b_{2n})
\widehat{f}_{\bar{T}}(\b_1',\cdots,\b_{2m}')
\label{c3}
\fin

Let us try to satisfy all this requirement. The simple form of the
left-right S-matrix allows to exclude it from the equations (\ref{a1},\ref{a2}).
Consider the function $g$ defined as follows:
\debut
f_{\Theta}(\b_1,\cdots,\b_{2n}\ |\ \b_1',\cdots,\b_{2m}')
=\prod\psi(\b_i,\b_j')g(\b_1,\cdots,\b_{2n}\ |\ \b_1',\cdots,\b_{2m}')
\label{fg}
\fin
where
$$\psi(\b,\b ')=
2^{-{3\over 4}}
\exp \(
-{1\over 4}(\b+\b ')
-\int _0^{\infty}
{2\sin ^2{1\over 2}(\b-\b ' +\pi i)k+\sinh ^2 {\pi k\over 2}
\over 2k \sinh \pi k \cosh {\pi k\over 2}}dk\)$$
The function $\psi (\b,\b ')$ satisfies the equations
\debut
&&\psi (\b,\b '+2\pi i)=\psi (\b,\b ')S_{RL}(\b '-\b),\quad
\psi (\b+2\pi i,\b ')=\psi (\b,\b ')S_{LR}(\b -\b ') \non\\ &&
\psi (\b,\b '+\pi i) \psi (\b,\b ')={1\over e^{\b}-i e^{\b '}},
\quad \psi (\b +\pi i,\b ') \psi (\b,\b ')={1\over ie^{\b}- e^{\b '}}
\label{psi}
\fin

It is clear that the equation (\ref{a2})
rewritten in terms of $g$
does not contain the left-right
S-matrices which means that the function $g$ must be of the form
\debut
g(\b_1,\cdots,\b_{2n}\ |\ \b_1',\cdots,\b_{2m}')=\sum\limits _{K,L}
c _{K,L}(\b_1,\cdots,\b_{2n}\ |\ \b_1',\cdots,\b_{2m}')
\widehat{f}^K(\b_1,\cdots,\b_{2n})\widehat{f}^L (\b_1',\cdots,\b_{2m}')
\label{xx}
\fin
where
$\widehat{f}^K(\b_1,\cdots,\b_{2n})$ and
$\widehat{f}^L (\b_1',\cdots,\b_{2m}')$ are different singlet solutions
(counted by $K$ and $L$ whose nature will be explained later)
of the equations
\debut
&&\widehat{f}^K(\cdots,\b_i,\b_{i+1},\cdots)S^Y(\b_i-\b_{i+1} )=
\widehat{f}^K(\cdots,\b_{i+1},\b_i,\cdots), \non\\ &&
\widehat{f}^K(\b_1,\cdots,\b_{2n-1},\b_{2n}+2\pi i)=
\widehat{f}^K(\b_{2n},\b_1,\cdots,\b_{2n-1}), \label{leq}\\ &&
\widehat{f}^L(\cdots,\b_i',\b_{i+1}',\cdots)S^Y(\b_i'-\b_{i+1}' )=
\widehat{f}^L(\cdots,\b_{i+1}',\b_i',\cdots), \non\\ &&
\widehat{f}^L(\b_1',\cdots,\b_{2m-1}',\b_{2m}'+2\pi i)=
\widehat{f}^L(\b_{2m}',\b_1',\cdots,\b_{2m-1}'), \non
\fin
The functions $c _{K,L}(\b_1,\cdots,\b_{2n}\ |\ \b_1',\cdots,\b_{2m}') $
are quasiconstants: $2\pi i$-periodical symmetrical with
respect to $ \b_1,\cdots,\b_{2n} $ and $ \b_1',\cdots,\b_{2m}'$
functions with possible singularities only at $\b_i,\b_i'=\pm\infty$.
The equations for left and right parts are the same, so, let us concentrate
for the moment only on the left one.

It is well known \cite{book,count} that the solutions to the equations
(\ref{leq}) are counted by the functions
$K(A_1,\cdots , $ $A_{n-1}|
B_1,\cdots ,B_{2n})$ which are antisymmetrical polynomials
of $A_1,\cdots ,A_{n-1}$ such that $1\le deg_{A_i}(K)\le 2n-1$,
$\forall i$
and symmetrical Laurent polynomials of $B_1,\cdots ,B_{2n} $. The solutions are
given by the formula
\debut
&&\widehat{f}^K(\b_1,\cdots ,\b_{2n})=d^n
\exp\({n\over 2}\sum \b _j\)\prod\limits _{i,j}\zeta(\b _i-\b _j)
\label{int} \\&& \times\int\limits _{-\infty}^{\infty}d\al _1\cdots
\int\limits _{-\infty}^{\infty}d\al _{n-1}
\prod\limits_{i,j}
\tilde{\varphi}(\al _i,\b _j)
\langle\Delta _n^{(0)}\rangle _n
(\al _1,\cdots ,\al _{n-1} |\b _{1},\cdots ,\b _{2n} )
K(e^{\al _1},\cdots ,e^{\al _{n-1}} |e^{\b _{1}},\cdots ,e^{\b _{2n}})
\non
\fin
where
$$
\tilde{\varphi}(\al _i,\b _j) =e^{-{1\over 2}(\al +\b)}
\Gamma \({1\over 4}+{\al -\b\over 2\pi i} \)
\Gamma \({1\over 4}-{\al -\b\over 2\pi i} \)
$$
We do not give here the formulae for
$\langle\Delta _n^{(0)}\rangle _n
(\al _1,\cdots ,\al _{n-1} |\b _{1},\cdots ,\b _{2n} ) $ which is a rational
function of all variables with values in the tensor product of
the isotopic spaces,  for $\zeta (\b)$ which is certain transcendental
function
and for the constant $d$:
these formulae can be found in the
book \cite{book} (Chapter 7).

It has to be emphasized that the integral (\ref{int})
vanishes for two kinds of function $K$ \cite{count,bbs2}:
\debut
&&K(A_1,\cdots ,A_{n-1}|B_1,\cdots ,B_{2n})=
\sum\limits _{k=1}^{n-1}(-1)^k (P(A_k)-P(-A_k))
K'(A_1,\cdots ,\widehat{A_k},\cdots ,A_{n-1}|B_1,\cdots ,B_{2n}) ,\non\\&&
K(A_1,\cdots ,A_{n-1}|B_1,\cdots ,B_{2n})=
\sum\limits _{k<l}(-1)^{k+l} C(A_k,A_l)
K''(A_1,\cdots ,\widehat{A_k},\cdots ,\widehat{A_l}\cdots ,A_{n-1}
|B_1,\cdots ,B_{2n})\quad \quad \label{zero}
\fin
where $K'$,$K''$ are some polynomials of the less number of variables
with the same properties as $K$,
$P(A)=\prod\limits _j (A_k+iB_j)$ and
$$
C(A_1,A_2 )={1\over A_1A_2}\left\{ {A_1-A_2\over A_1+A_2 }
(P(A_1)P(A_2)-P(-A_1)P(-A_2))
+
(P(-A_1)P(A_2)-P(A_1)P(-A_2))\right\}  \label{C}
$$
So, the polynomials $K$ are defined modulo the polynomials
of the kind (\ref{zero}) , the fact that has been used in \cite{count}
to show that we have correct number of solutions to (\ref{leq}).

Combining (\ref{fg}),(\ref{xx}) and (\ref{int}) we find that
the from factors satisfying (\ref{a1}) and (\ref{a2}) are of the form
\debut
&&f_{\Theta}(\b_1,\cdots,\b_{2n}\ |\ \b_1',\cdots,\b_{2m}')
=M^2\prod\psi(\b_i,\b_j')
\prod\limits _{i,j}\zeta(\b _i-\b _j)
\prod\limits _{i,j}\zeta(\b _i'-\b _j')
\non\\ &&\times
\int\limits _{-\infty}^{\infty}d\al _1\cdots
\int\limits _{-\infty}^{\infty}d\al _{n-1}
\int\limits _{-\infty}^{\infty}d\al '_1\cdots
\int\limits _{-\infty}^{\infty}d\al '_{m-1}
\prod\limits _{i=1}^{n-1}\prod\limits_{j=1}^{2n} \tilde{\varphi}(\al _i,\b _j)
\prod\limits _{i=1}^{m-1}\prod\limits_{j=1}^{2m} \tilde{\varphi}(\al '_i,\b '_j)
\non\\ &&\times
\langle\Delta _n^{(0)}\rangle _n
(\al _1,\cdots ,\al _{n-1} |\b _{1},\cdots ,\b _{2n} )
\langle\Delta _n^{(0)}\rangle _n
(\al' _1,\cdots ,\al _{m-1}' |\b _{1}',\cdots ,\b _{2m}' )
\non\\ &&\times
M_{n,m}(e^{\al _1},\cdots ,e^{\al _{n-1}} |
e^{\al _1'},\cdots ,e^{\al _{m-1}'}|
e^{\b _{1}},\cdots ,e^{\b _{2n}}|
e^{\b _{1}'},\cdots ,e^{\b _{2m}'})
\label{ff}
\fin
where
$ M_{n,m}(A _1,\cdots ,A _{n-1} |
A _1',\cdots ,A _{m-1}'|
B _{1},\cdots ,B _{2n}|
B _{1}',\cdots ,B _{2m}')  $
is an antisymmetrical polynomial of $ A _1,$  $\cdots ,A _{n-1} $
($A _1',\cdots ,A _{m-1}'$) whose degree with respect to every
variable is from $1$ to $2n-1$ (from $1$ to $2m-1$) and
symmetrical Laurent polynomial of $B _{1},\cdots ,B _{2n}$
($B _{1}',\cdots ,B _{2m}'$).

Now we have to satisfy the rest of our requirements on the form
factors. In the paper \cite{count} there is a general prescription
for handling the residue condition (\ref{a3}) for the integrals of the
form (\ref{int}). Applying this prescription to our situation and
using the equations (\ref{psi}) one finds that the residue condition
(\ref{a3}) is satisfied if and only if the function $M_{n,m}$ possesses the
properties:
\newline
First,
\debut
&&M_{n,m}(A _1,\cdots ,A _{n-1} |A _1',\cdots ,A _{m-1}'|
B _{1},\cdots ,B _{2n-2},B,-B|B _{1}',\cdots ,B _{2m}')=\non\\&&=
\sum\limits _{k=1}^{n-1}(-1)^k\prod\limits _{p\ne k}(A_p^2+B^2)
M_{n-1,m}^k(A _1,\cdots ,A _{n-1} |A _1',\cdots ,A _{m-1}'|
B _{1},\cdots ,B _{2n-2}|B|B _{1}',\cdots ,B _{2m}') \non\\  &&
M_{n,m}(A _1,\cdots ,A _{n-1} |A _1',\cdots ,A _{m-1}'|
B _{1},\cdots ,B _{2n}|B _{1}',\cdots ,B _{2m-2}',B',-B')= \label{rel1}\\&&=
\sum\limits _{k=1}^{m-1}(-1)^k\prod\limits _{p\ne k}((A_p')^2+(B')^2)
M_{n,m-1}^k(A _1,\cdots ,A _{n-1} |A _1',\cdots ,A _{m-1}'|
B _{1},\cdots ,B _{2n}|B _{1}',\cdots ,B _{2m-2}'|B')
\non
\fin
where $M_{n-1,m}^k $ and $M_{n,m-1}^k$
are some {\it polynomials} in $A_i$ and
$A_i'$.
\newline
Second,
\debut
&&M_{n-1,m}^k(A _1,\cdots ,A _{k-1},\pm iB ,A _{k-1},\cdots ,A _{n-1}
|A _1',\cdots ,A _{m-1}'|
B _{1},\cdots ,B _{2n-2}|B|B _{1}',\cdots ,B _{2m}')=\non \\&& =
\pm B\prod\limits _{j=1}^{2m}(B\mp iB'_j)
M_{n-1,m}(A _1,\cdots ,A _{k-1},A _{k-1},\cdots ,A _{n-1}
|A _1',\cdots ,A _{m-1}'|
B _{1},\cdots ,B _{2n-2}|B _{1}',\cdots ,B _{2m}'), \non\\&&
M_{n,m-1}^k(A _1,\cdots ,A _{n-1}
|A _1',\cdots ,A _{k-1}',\pm iB' ,A _{k-1}',\cdots ,A _{m-1}'|
B _{1},\cdots ,B _{2n}|B _{1}',\cdots ,B _{2m-2}'|B')=
 \label{rel2} \\ &&=
\pm B'\prod\limits _{j=1}^{2m}(B'\pm iB_j)
M_{n,m-1}(A _1,\cdots ,A _{n-1}
|A _1',\cdots ,A _{k-1}',A _{k-1}',\cdots ,A _{m-1}'|
B _{1},\cdots ,B _{2n}|B _{1}',\cdots ,B _{2m-2}')\non
\fin
These equations are necessary and sufficient for the formula
(\ref{ff}) to define form factors of a local operator. Certainly,
they have infinitely many solutions. We shall give only one of these solutions
corresponding to the operator $\Theta$. Let us introduce the notations
for the sets of integers:
$S=\{1,\cdots ,2n\}$, $S'=\{1,\cdots ,2m\}$ then
\debut
&&M_{n,m}(A _1,\cdots ,A _{n-1} |A _1',\cdots ,A _{m-1}'|
B _{1},\cdots ,B _{2n}|B _{1}',\cdots ,B _{2m}') =
\prod\limits _{i<j}(A_i-A_j)\prod\limits _{i<j}(A_i'-A_j')
\non\\&& \times \prod\limits _{j=1}^{2n} B_j^{-1}
\prod\limits _{i=1}^{n-1} A_i^2
\prod\limits _{i=1}^{m-1} A_i '
\sum\limits _{{T\subset S\atop\# T=n-1}}
\sum\limits _{{T'\subset S'\atop\# T'=m-1}}
\prod\limits _{j\in T} B_j
\prod\limits _{i=1}^{n-1}
\prod\limits _{j\in T}(A_i+iB_j)
\prod\limits _{i=1}^{m-1}
\prod\limits _{j\in T'}(A_i'+iB_j') \non\\&&
\times
\prod\limits _{{i,j\in \bar{T}\atop i<j}}(B_i+B_j)
\prod\limits _{{i,j\in \bar{T}'\atop i<j}}(B_i'+B_j')
\prod\limits _{{i\in T\atop j\in \bar{T}}}{1\over B_i-B_j}
\prod\limits _{{i\in T'\atop j\in \bar{T}'}}{1\over B_i'-B_j'} \non\\&& \times
\prod\limits _{{i\in T\atop j\in T'}}(B_i+iB_j')
\prod\limits _{{i\in \bar{T}\atop j\in \bar{T}'}}(B_i-iB_j')
\ X_{T,T'}(B_1,\cdots ,B_{2n}|B_1',\cdots ,B_{2m}')
\label{poly}
\fin
where $\bar{T}=S\backslash T$,  $\bar{T'}=S'\backslash T'$,
\debut
X_{T,T'}(B_1,\cdots ,B_{2n}|B_1',\cdots ,B_{2m}')=
\sum\limits _{i_1,i_2\in\bar{T}}
\prod\limits _{p=1}^2\(
{\prod\limits _{j\in T}(B_{i_p}+B_j)
\over  \prod\limits _{j\in \bar{T}\backslash \{i_1,i_2\}}(B_{i_p}-B_j)}
{\prod\limits _{j\in T'}(B_{i_p}+iB_j')
\over  \prod\limits _{j\in \bar{T}'}(B_{i_p}-iB_j')} \)
\non
\fin
The polynomial $X_{T,T'}(B_1,\cdots ,B_{2n}|B_1',\cdots ,B_{2m}') $
is in fact quite symmetric with respect to replacement
$B\leftrightarrow B'$.

Let us show that $M_{n,m}$ satisfies all necessary requirements.
The relations (\ref{rel1},\ref{rel2}) are easily checked using the
formula
$$\prod\limits _{i=1}^{n-1}
\prod\limits _{j=1}^{n-1}(A_i+iB_j) \prod\limits _{i<j}(A_i-A_j)=
\prod\limits _{i=1}^{n-1}
\prod\limits _{j=1}^{n-1}(A_i^2+B_j^2)
\prod\limits _{i<j}^{n-1}{1\over B_i-B_j}\ det\({1\over A_i-iB_j}\) $$
So, $M_{n,m}$  really defines a local operators. We have to
show that the additional conditions formulated at the
beginning of this section are satisfied in order to show
that this local operator is indeed the trace of the energy-momentum
tensor.

Obviously, $M_{n,m}$ is a homogeneous function of all its
variables ($A,B,A',B'$) of total degree $(m+n)^2-2m-2n$.
Considering the formula (\ref{ff}) one realizes that this fact
provides that the operator defined by $M_{n,m}$ is Lorentz
scalar i.e. its form factors are invariant under simultaneous
shift of all the rapidities.
Let us consider now the conditions 1-3 formulated earlier.

We start form the condition 3. One finds that
$$\psi (\b +\log \Lambda ,\b ')\to \Lambda ^{-{1\over 2}} e^{-{1\over 2}\b} $$
The integrals with respect to $\al _i$ in (\ref{ff}) are
concentrated near the points $\b _j$, so when $\b_j$ become of
order $\log \Lambda $ the integration variables $\al _i$ must be of the
same order. One finds that when  $\log \Lambda \to \infty$
\debut
&&M_{n,m}(\Lambda A _1,\cdots ,\Lambda A _{n-1} |A _1',\cdots ,A _{m-1}'|
\Lambda B _{1},\cdots ,\Lambda B _{2n}|B _{1}',\cdots ,B _{2m}') \to\non\\
&&\to
\Lambda ^{2mn+n^2-2n-2}
\prod\limits _{j=1}^{2n} B_j^{2m-1}
\sum\limits _{j=1}^{2n} B_j^{-1}
\prod\limits _{i=1}^{n-1} A_i^3
\prod\limits _{i<j}(A_i^2-A_j^2)
\sum\limits _{j=1}^{2m} B_j '
\prod\limits _{i=1}^{m-1} A_i '
\prod\limits _{i<j}(A_i'^2-A_j'^2)
\non
\fin
This formula is equivalent to (\ref{c3}) because
the form factors of the energy-momentum tensor of $SU(2)$-invariant
Thirring model $\widehat{f}_{T}(\b_1,\cdots,\b_{2n})$
$\widehat{f}_{\bar{T}}(\b_1',\cdots,\b_{2m}') $ are given by the formulae
on the type (\ref{int}) with the polynomial $K$ equal respectively to
\debut
M^2\prod\limits _{j=1}^{2n} B_j^{-1}
\sum\limits _{j=1}^{2n} B_j^{-1}\prod\limits _{i=1}^{n-1} A_i^3
\prod\limits _{i<j}(A_i^2-A_j^2)
\qquad and
\qquad M^2\sum\limits _{j=1}^{2m} B_j '
\prod\limits _{i=1}^{m-1} A_i '
\prod\limits _{i<j}(A_i'^2-A_j'^2)
\label{T}
\fin

Let us consider the condition (\ref{c2}).
One finds that
\debut
&&\left.
{1\over B _{1}' +B _{2}' }
M_{n,1}(A _1,\cdots ,A _{n-1} |\emptyset|
B _{1},\cdots ,B _{2n}|B _{1}' ,B _{2}') \right| _{B _{2}' =-B _{1}'}=
\non\\&&\hskip 1cm =
\prod\limits _{j=1}^{2n} B_j^{-1}
\prod\limits _{i=1}^{n-1} A_i^3
\prod\limits _{i<j}(A_i^2-A_j^2)
\(\prod (B_1'+iB_j )-\prod (B_1'-iB_j )\)
\non
\fin
which together with (\ref{T}) gives the first equation from (\ref{c2}),
the second relation is proven similarly.

The condition (\ref{c1}) is the most complicated to prove.
Naively it has to be equivalent to the fact that $M_{n,m} $
is divisible by $\sum B_j^{-1}$ and $\sum B_j'$, but that is not
the case: the function $M_{n,m} $ has to be substituted into
the integral hence it is defined modulo the functions of
the type (\ref{zero}) (and similar functions of $A_i'$). Thus
the divisibility has to be proven modulo these null-polynomials.
We have checked this fact for many particular examples, but still
we lack a general proof. However, the calculations in particular
cases go so nicely that we have no doubt that the relation (\ref{c1})
is satisfied generally.

\section{Some generalizations.}

The model considered in this paper provides a special
case of wide class of massless flows. Consider the massless
flow \cite{sot} between the UV
coset model $su(2)_{k+1}\otimes su(2)_k/su(2)_{2k+1}$ and
the IR coset model $su(2)_{k}\otimes su(2)_1/su(2)_{k+1}$, 
the latter model is nothing but  the minimal
model $M_{k+2}$. This flow is defined in UV by the relevant
operator of dimension 
$1-2/(2k+3) $, 
it arrives at IR region along $T\bar{T}$.
The massless S-matrices for these flows are written in terms of
RSOS restriction of the sine-Gordon (SG) S-matrix $S^{\xi}(\b)$
($\xi$ is SG coupling constant defined as in \cite{book}).
Namely \cite{bl},
$$S_{LL}(\b _1,\b_2)=S^{\pi(k+2)}_{RSOS}(\b _1-\b_2),\qquad
S_{RR}(\b _1',\b_2')=S^{\pi(k+2)}_{RSOS}(\b _1'-\b_2') $$
The left-right S-matrix is independent of $k$, it is the same as above.
When $k=\infty$ the model coincides with PCM${}_1$. Another
extreme case is $k=1$ when the model describes the flow
between tricritical and critical Ising models. It is well known
that the RSOS-restriction for $\xi =3\pi$ effectively
reduces soliton to one-component particle with free scattering:
$$S^{\pi(k+2)}_{RSOS}(\b)=-1$$

The results of this paper allow straightforward generalization to
these flows. One has to replace the formulae of the type (\ref{int}) by
their SG analogues. This does not disturb the function
$M_{n,m}$ because the way of counting solution to the equation
of the type (\ref{leq}) in SG case does not depend on the
coupling constant as well as all the equations on $M_{n,m}$.
So, the form factors are defined by (\ref{ff}) where one has to
replace the
functions $\zeta$, $\varphi$, $\langle \Delta ^{(0)}\rangle$
by their SG-analogues and to take RSOS restriction.

Let us see how it works in the case $k=1$. For generic
coupling constant the
formulae (\ref{zero}) present the only reason for vanishing the
integrals of the type
(\ref{int}). However when $ \xi=3\pi$ and RSOS restriction is
taken the integral does not vanish only for the antisymmetrical
polynomial $K(A_1,\cdots ,A_{n-1})$ of very special kind:
$$ K(A_1,\cdots ,A_{n-1}) =\prod A_i^2\prod\limits _{i<j}(A_i^2-A_j^2)$$
The value of the integral for this kind of polynomial
(taking in account the functions $\zeta$ also) is
$$\prod\limits _{i<j}\tanh {1\over 2}(\b_i-\b _j)
\exp\({1\over 2}\sum \b_j \)$$
Consider the formula (\ref{poly}). We have to take the functions
$$
\prod\limits _{i<j}(A_i-A_j)
\prod\limits _{i=1}^{n-1} A_i^2
\prod\limits _{i=1}^{n-1}
\prod\limits _{j\in T}(A_i+iB_j)
\quad and\quad
\prod\limits _{i<j}(A_i'-A_j')
\prod\limits _{i=1}^{m-1} A_i '
\prod\limits _{i=1}^{m-1}
\prod\limits _{j\in T'}(A_i'+iB_j'),
$$
to decompose them with respect to antisymmetrical polynomials
of $A_i$ and $A'_i$ corresponding to different partitions
and to find the coefficients with which enter
the polynomials $\prod A_i^2\prod\limits _{i<j}(A_i^2-A_j^2)$ and
$\prod (A'_i)^2\prod\limits _{i<j}((A'_i)^2-(A'_j)^2)$. These coefficients are
$$
\prod\limits _{j\in T}B_j
\prod\limits _{{i,j\in T\atop i<j}}(B_i+B_j)
\quad and \quad
\prod\limits _{{i,j\in T'\atop i<j}}(B_i'+B_j') $$
Thus we find the following formula for the form factors of $\Theta$ for
this model
\debut
&&f_{\Theta}(\b_1,\cdots,\b_{2n}\ |\ \b_1',\cdots,\b_{2m}')
=M^2 \prod\psi(\b_i,\b_j')
\prod\limits _{i,j}\tanh {1\over 2}(\b_i-\b _j)
\prod\limits _{i,j}\tanh {1\over 2}(\b_i-\b _j)
\non\\ &&\times
\exp\({1\over 2}\sum \b_j+{1\over 2}\sum \b_j' \)
Q_{n,m}
(e^{\b _{1}},\cdots ,e^{\b _{2n}}|
e^{\b _{1}'},\cdots ,e^{\b _{2m}'})
\non
\fin
where
\debut
&&Q_{n,m}(
B _{1},\cdots ,B _{2n}|B _{1}',\cdots ,B _{2m}') =
\non\\&& = \prod\limits _{j=1}^{2n} B_j^{-1}
\sum\limits _{{T\subset S\atop\# T=n-1}}
\sum\limits _{{T'\subset S'\atop\# T'=m-1}}
\prod\limits _{j\in T} B_j^2
\prod\limits _{{i,j\in T\atop i<j}}(B_i+B_j)
\prod\limits _{{i,j\in T'\atop i<j}}(B_i'+B_j')
\non\\&&
\times
\prod\limits _{{i,j\in \bar{T}\atop i<j}}(B_i+B_j)
\prod\limits _{{i,j\in \bar{T}'\atop i<j}}(B_i'+B_j')
\prod\limits _{{i\in T\atop j\in \bar{T}}}{1\over B_i-B_j}
\prod\limits _{{i\in T'\atop j\in \bar{T}'}}{1\over B_i'-B_j'} \non\\&& \times
\prod\limits _{{i\in T\atop j\in T'}}(B_i+iB_j')
\prod\limits _{{i\in \bar{T}\atop j\in \bar{T}'}}(B_i-iB_j')
X_{T,T'}(B_1,\cdots ,B_{2n}|B_1',\cdots ,B_{2m}')
\label{kon}
\fin
one can write a formula for this polynomial in determinant form,
but we think that (\ref{kon}) shows quite transparently how all the
required properties of this polynomial \cite{muss} are satisfied.

\end{document}